\documentclass{article}
\usepackage{spconf,amsmath,amssymb,graphicx,booktabs}
\usepackage[hidelinks]{hyperref}
\usepackage{balance}
\usepackage{xcolor}
\newcommand{\CRBRangeTen}{0.243}
\newcommand{\CRBVelTen}{0.00367}
\newcommand{\TCREIdealRangeTen}{0.246}
\newcommand{\TCREIdealVelTen}{0.00373}

\newcommand{\ZFFTTenRange}{1.289}
\newcommand{\TCRETenRange}{0.702}
\newcommand{\FullPhaseTenRange}{0.671}

\newcommand{\ZFFTTenVel}{0.02924}
\newcommand{\TCRETenVel}{0.00910}

\newcommand{\RangeRedMin}{32.8}
\newcommand{\RangeRedMax}{46.3}
\newcommand{\VelRedMin}{58.6}
\newcommand{\VelRedMax}{71.0}
\newcommand{\RangeRedTen}{45.6}
\newcommand{\RangeCILowTen}{36.2}
\newcommand{\RangeCIHighTen}{51.9}
\newcommand{\VelRedTen}{68.9}
\newcommand{\VelCILowTen}{63.7}
\newcommand{\VelCIHighTen}{74.2}
\newcommand{\BlockTheoryPenaltyPct}{3.27}
\newcommand{\BlockEmpiricalPenaltyPct}{2.13}
\newcommand{\CommonSlopeTwelve}{0.00371}
\newcommand{\FullChirpTwelve}{1.495}
\newcommand{\CommonSlopeRatioTwelve}{403}
\newcommand{\CaptureNoiselessM}{0.600}
\newcommand{\CaptureSubmmSampledM}{0.525}
\newcommand{\MotionDistanceMaxMm}{39.69}
\newcommand{\MotionNarrowVel}{0.307}
\newcommand{\MotionCoupledVel}{0.00373}
\newcommand{\MotionCoupledRange}{0.267}
\newcommand{\SequentialUpdates}{11920}
\newcommand{\SequentialBoundaryViolations}{0}
\newcommand{\SequentialBranchFailures}{0}
\newcommand{\SequentialTrackLosses}{0}

\newcommand{\SequentialMaxPredRangeMm}{5.48}
\newcommand{\SequentialMaxPredVelCms}{1.87}

\newcommand{\CN}{\mathcal{CN}}

\title{TRACK-CONDITIONED RESIDUAL FREQUENCY ESTIMATION FOR LOW-RESOLUTION FMCW RADAR}
\name{Huy Trinh \qquad George Shaker}
\address{
Department of Electrical and Computer Engineering,
University of Waterloo,
Canada\\
h3trinh@uwaterloo.ca \qquad gshaker@uwaterloo.ca
}

\begin{document}
\maketitle

\begin{abstract}
Low-resolution FMCW radar normally forms a complete range--Doppler representation before detection and tracking. After association, however,
the tracker already predicts the target's next range and radial velocity.
We propose track-conditioned residual estimation (TCRE), which removes the
predicted moving-target phase before spectral formation, summarizes each
64-sample chirp by four coherent complex sums, and estimates the remaining
beat- and Doppler-frequency errors from phase progression. We account for
the first-order fast/slow-time coupling of a moving target and derive the
local phase-unwrapping region and three-receiver local information bound.
Four summaries retain 93.77\% of the zero-residual fast-time information
with a $16\times$ smaller retained representation. In 5,000 matched
60-GHz cluttered simulations, TCRE reduces range and velocity RMSE by
\RangeRedMin--\RangeRedMax\% and \VelRedMin--\VelRedMax\%, respectively,
relative to a fully specified same-prior residual ZFFT.
\end{abstract}
\section{Introduction}
\label{sec:intro}
Frequency-modulated continuous-wave (FMCW) radar obtains range and radial velocity from compact coherent measurements using standard fast- and slow-time processing in various applications~\cite{richards,patole,trinh2026reliablequasistaticpostfallflooroccupancy,trinh2026physicsinformeddigitaltwinframework,trinh2026lightweightrangeangleimagingbased}. A broad range--Doppler (RD) search is useful during acquisition since the target state is unknown. However, after a target has been detected and associated, the tracker already predicts where its return should appear in the next frame. Thus, repeating the same broad search before using the prediction can therefore become less efficient in the tracked mode. 

Our previous work on the same low-resolution 60-GHz radar class also studied the representation design for the problem of quasi-static occupancy detection~\cite{trinh_rasso}. In that work, Doppler-domain warping was applied \emph{after} RD formation to allocate more samples around near-zero Doppler. Here, we use prior information earlier in the chain. As shown in~Fig.~\ref{fig:flow}, the predicted range and velocity are applied directly to coherent 
IF samples \emph{before} a new spectrum is formed, translating the absolute range/Doppler problem to a close-to-zero residual-frequency problem. We therefore propose track-conditioned residual estimation (TCRE). TCRE
removes the predicted range/Doppler phase, retains a few coherent sums
per chirp, and refines residual range and velocity using standard
phase-regression ideas. The main contributions of our research can be summarized in two points:
 i) a coupling-aware coherent-block formulation with an explicit capture region and retained-information analysis; and ii) controlled known-truth evaluation against strengthened global FFT, fully specified same-prior residual ZFFT, and an uncompressed phase-estimation control, including moving-model, block-count, prior-error, capture-boundary, and sequential tests.
\section{Related Work}
\label{sec:related}
The residual model used by TCRE is built on classical single-tone frequency estimation. 
Rife and Boorstyn derived maximum-likelihood estimators and the corresponding Cram\'er--Rao bounds (CRBs) for discrete-time tones~\cite{rife}.
In~\cite{tretter}, Tretter showed that at sufficiently high SNR, the noisy phase can be treated as an approximately linear regression problem. Kay \emph{et al.}~\cite{kay} developed an efficient weighted single-frequency estimator. 
Local FMCW refinement is also well established. Wang \emph{et al.} refine rough range estimates using a local resampling Fourier transform \cite{wang_lrft}. Neemat \emph{et al.}~\cite{neemat} study reconfigurable range--Doppler processing. 
Moussa and Liu analyze wideband FMCW range--Doppler estimation
and the motion-dependent terms that arise beyond the conventional
separable model~\cite{moussa}. Dai \emph{et al.}~\cite{dai} derive high-resolution beat-frequency and phase estimation and compare it with a CRB. A tracked receiver could similarly center a ZFFT around the predicted state. 

Recent event-driven approaches reduce dense spectral processing in a different way.
Guo \emph{et al.} formulate asynchronous spectral estimation from event
measurements using Event-Driven Prony~\cite{guo}, and
Pavlicek \emph{et al.} extend this direction toward radar spectral
super-resolution and hardware \cite{pavlicek}. Reeb \emph{et al.} use
spiking neural resonators to process FMCW samples for range--angle
estimation \cite{reeb}, while Chiavazza \emph{et al.} develop
chirp-wise spike-based range--velocity processing to avoid waiting for
a complete RD frame \cite{chiavazza}. TCRE is complementary to these
approaches: it assumes conventional coherent samples and an existing tracker rather than introducing a new event encoder or acquisition mechanism.
\section{Methodology}
\label{sec:method}
\subsection{Radar configuration and tracked-mode signal}
\begin{figure}[t]
\centering
\includegraphics[width=0.90\columnwidth]{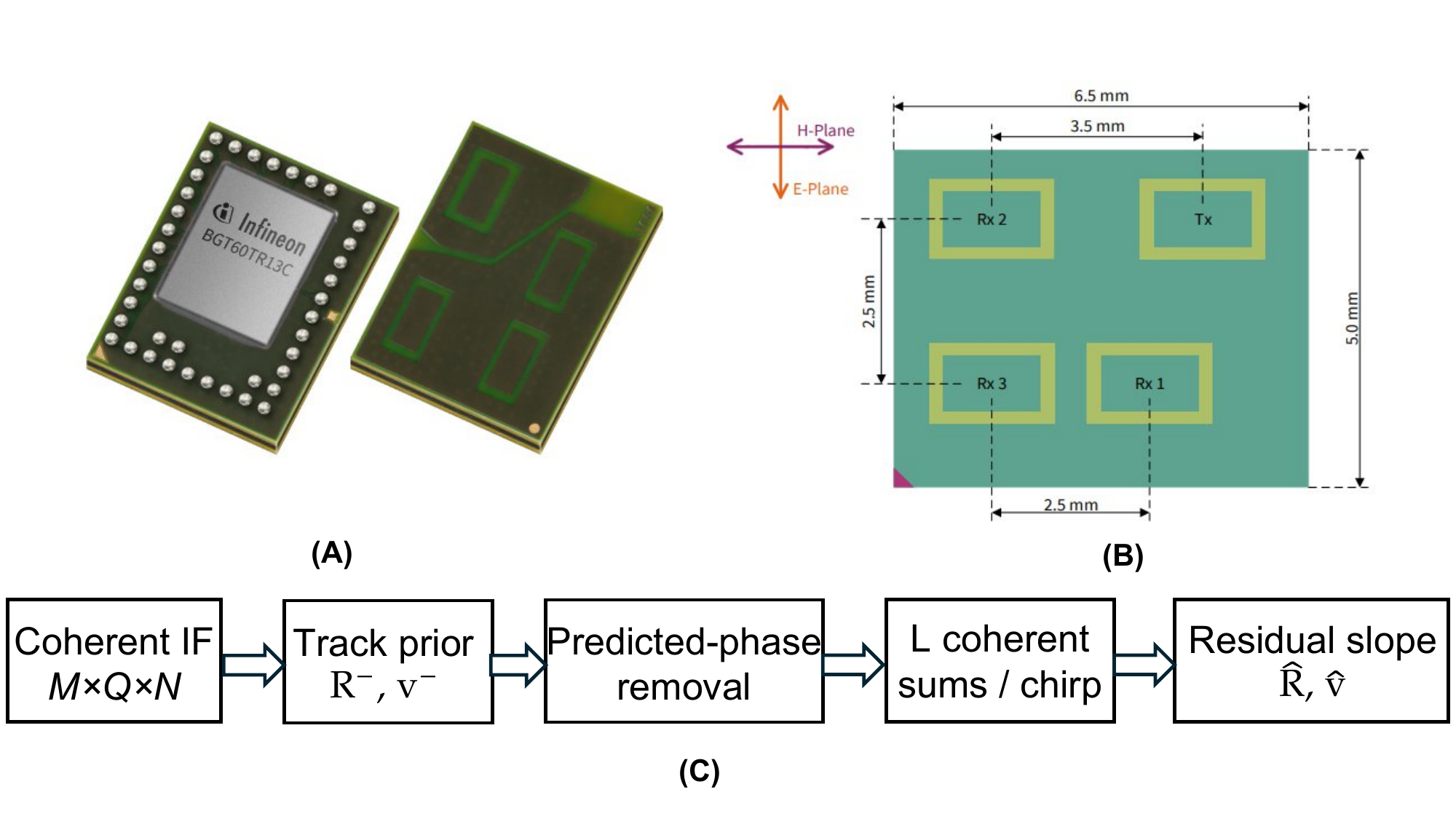}
\caption{Radar reference and tracked-mode processing: \textbf{(a)} BGT60TR13C package, \textbf{(b)} 1-TX/3-RX geometry, and \textbf{(c)} TCRE flow. FFT processing remains available for acquisition/reacquisition and TCRE starts after the tracker supplies $(R^-,v^-)$.}
\label{fig:flow}
\end{figure}
Figure~\ref{fig:flow}(a,b) shows the BGT60TR13C-class hardware and 1-TX/3-RX geometry used as the platform reference~\cite{trinh_rasso,infineon}. The angle is not estimated here and the three receiver channels are retained because their independent noise contributes information, while receiver-dependent phases are nuisance terms.

Let $m$, $q$, and $n$ denote receiver, chirp, and fast-time indices.
With $T_s=1/f_s$, $T_c=NT_s$, $S=B/T_c$, and
$\lambda=c/f_c$, define $t_n=nT_s$ and
$\tau_q=(q-(Q-1)/2)T_r$. Range $R$ is referenced to
$\tau_q=t_n=0$, the temporal midpoint between the two central chirp
starts for even $Q$. Following the moving-target FMCW model in
\cite{moussa}, the first-order deramped signal is
\begin{equation}
\begin{split}
 x_m[q,n]=a_m\exp\{-j2\pi[&f_D\tau_q+f_b t_n\\&+\kappa t_n\tau_q\}+w_m[q,n],
 \end{split}
 \label{eq:model_coupled}
\end{equation}
where
$f_D=2v/\lambda$, $f_b=2SR/c+f_D$, and
$\kappa=2Sv/c=(S/f_c)f_D$. The complex coefficient $a_m$
contains target amplitude, initial phase, and receiver-dependent
constant phase, while
$w_m[q,n]\sim\CN(0,\sigma^2)$. SNR is the per-receiver,
per-fast-time-sample value
$\gamma=|a_m|^2/\sigma^2$. We then set $|a_m|=1$ for all
$M=3$ receivers with independent noise.
After acquisition and association, the tracker supplies $(R^-,v^-)$
and hence $(f_b^-,f_D^-,\kappa^-)$. We assume one dominant component
within the local track. In addition, overlapping multi-target returns are outside this present paper's scope. TCRE removes the complete first-order predicted
phase,
\begin{equation}
\begin{split}
 z_m[q,n]&=x_m[q,n]e^{j2\pi[f_D^-\tau_q+f_b^-t_n+\kappa^-t_n\tau_q]}\\
 &\approx a_m e^{-j2\pi[\Delta f_D\tau_q+\Delta f_b t_n+\Delta\kappa t_n\tau_q]}+\eta_m[q,n],
\end{split}
\label{eq:residual_coupled}
\end{equation}
where $\Delta f_b=f_b-f_b^-$, $\Delta f_D=f_D-f_D^-$, and $\Delta\kappa=(S/f_c)\Delta f_D$. Therefore, a good prediction moves the target near zero residual frequency, leaving only the prediction error to be estimated.
\subsection{Proposed TCRE processing}
Rather than forming another local spectrum, we preserve a small number of coherent observations on which classical phase estimation can operate. Each $N$-sample chirp is divided into $L$ blocks of $P=N/L$ samples, block-center time
$t_\ell=[\ell P+(P-1)/2]T_s$. 
Moving-target coupling makes the fast residual weakly chirp-dependent, $g_q=\Delta f_b+\frac{S}{f_c}\Delta f_D\tau_q,$
and coherently summing the $P$ samples in each block gives
\begin{equation}
 b_m[q,\ell]=a_mD_P(g_q)
 e^{-j2\pi[\Delta f_D\tau_q+g_qt_\ell]}+\nu_m[q,\ell].
 \label{eq:block_coupled}
\end{equation}
Here
$D_P(f)=\sin(\pi P fT_s)/\sin(\pi fT_s)$ is the length-$P$
coherent-sum response and
$\nu_m[q,\ell]\sim\CN(0,P\sigma^2)$.
This expression explains why summarization does not simply discard the desired state. In detail, the block then provides coherent gain near the prediction, while the phase progression across chirps and block centers still carries the residual Doppler and beat frequencies. Following the high-SNR phase-regression interpretation in references~\cite{tretter,kay}, we unwrap the block phases and use magnitude-squared weighting 
\begin{equation}
\begin{split}
\phi_m[q,\ell]\approx\beta_{m\ell}-2\pi\{&\Delta f_D(1+St_\ell/f_c)\tau_q\\
 &+\Delta f_b t_\ell\}+\epsilon_m[q,\ell].
\end{split}
\label{eq:phase_coupled}
\end{equation}
We fit one common $\Delta f_D$ using the effective slow-time coordinate
$\tau_q(1+St_\ell/f_c)$ while retaining one nuisance intercept for every
receiver/block pair. After removing the estimated Doppler term and its
coupling, $\Delta f_b$ is estimated from phase progression across
$t_\ell$. 
Using the standard FMCW range and Doppler relationships, the state
update is
$\hat v=v^-+\frac{\lambda}{2}\widehat{\Delta f_D}$ and $\hat R=R^-+\frac{c}{2S} (\widehat{\Delta f_b}-\widehat{\Delta f_D})$.
The tracked update therefore retains $MQL$ complex block values rather than $MQN$ fast-time samples.
\subsection{Locality and retained information}
\begin{figure}[t]
\centering
\includegraphics[width=.49\columnwidth]{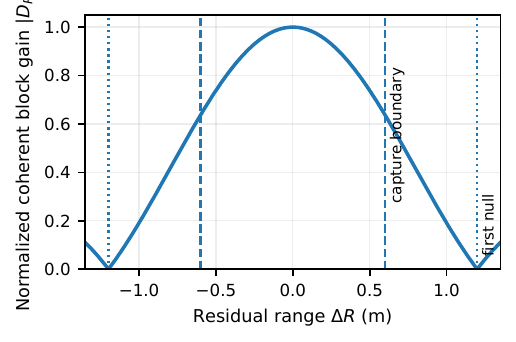}\hfill
\includegraphics[width=.49\columnwidth]{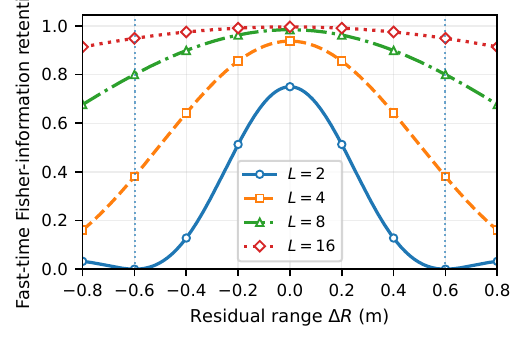}
\caption{Local effect of coherent summarization. \textbf{Left}: normalized $P=16$ block response. The dashed lines show the adjacent-block phase boundary, and dotted lines the first null. \textbf{Right}: analytical fast-time Fisher-information retention for several block counts. The tracker places normal operation near $\Delta R=0$.}
\label{fig:theory}
\end{figure}

Because coupling makes the effective fast residual chirp dependent,the local phase-unwrapping condition is
\begin{equation}
\begin{aligned}
 \max_q\left|\Delta f_b+\frac{S}{f_c}\Delta f_D\tau_q\right|&<\frac{1}{2PT_s},\\
 \max_\ell\left|\Delta f_D(1+St_\ell/f_c)\right|&<\frac{1}{2T_r}.
\end{aligned}
\label{eq:capture_corrected}
\end{equation}
For $\Delta f_D=0$ and $L=4$, the first condition gives us the noiseless $\pm\CaptureNoiselessM$-m range boundary. From the same finite-sum model and the classical single-tone CRB time-spread argument \cite{rife}, the zero-residual fast-time information fraction is $ \eta_b(0)=\frac{N^2-P^2}{N^2-1}=0.9377\quad(N=64,L=4),$
so four sums retain 93.77\% of the local fast-time information with a $16\times$ representation reduction as shown in Fig.~\ref{fig:theory}. With $M$ independent receivers,
and receiver/block phase intercepts treated as nuisance parameters, the 
local slope bounds are
\begin{align}
 \operatorname{var}(\widehat{\Delta f_b}) &\ge
 \frac{1}{2MP\gamma(2\pi)^2Q\sum_\ell(t_\ell-\bar t)^2},\\
 \operatorname{var}(\widehat{\Delta f_D}) &\ge
 \frac{1}{2MP\gamma(2\pi)^2
 [\sum_\ell(1+St_\ell/f_c)^2][\sum_q\tau_q^2]}.
 \label{eq:crb_multi_rx}
\end{align}
where $\bar t=L^{-1}\sum_{\ell=0}^{L-1}t_\ell$. Using $R=c(f_b-f_D)/(2S)$ and $v=\lambda f_D/2$, the 10-dB $M=3$,
$L=4$ bounds are \CRBRangeTen\,mm and \CRBVelTen\,cm/s. A 
5,000-trial 
moving-signal experiment gives
\TCREIdealRangeTen\,mm and \TCREIdealVelTen\,cm/s, respectively.
\subsection{Known-truth simulation and baselines}
\begin{table}[t]
\centering
\caption{Low-resolution FMCW simulation configuration.}
\label{tab:setup}
\setlength{\tabcolsep}{3.2pt}
\begin{tabular}{lc@{\qquad}lc}
\toprule
Parameter & Value & Parameter & Value\\
\midrule
$f_c$ & 60 GHz & $B$ & 500 MHz\\
$T_c$ & 64 $\mu$s & $f_s$ & 1 MHz\\
$N$ & 64 & $Q$ & 128\\
$T_r$ & 0.4167 ms & $M$ & 3\\
$L$ & 4 & $P=N/L$ & 16\\
$\sigma_R$ & 3 cm & $\sigma_v$ & 6 cm/s\\
SNR & 0--20 dB & & \\
\bottomrule
\end{tabular}
\end{table}
\begin{figure}[t]
\centering
\includegraphics[width=.49\columnwidth]{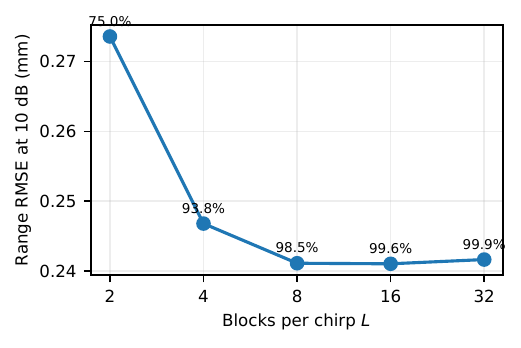}\hfill
\includegraphics[width=.49\columnwidth]{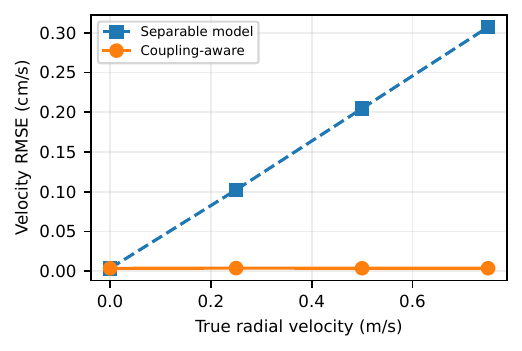}
\caption{\textbf{Left:} 10-dB block-count tradeoff with retained-information labels. \textbf{Right:} Exact moving-target validation; omitting fast/slow-time coupling causes velocity error to grow with target speed.}
\label{fig:blocktrade}
\end{figure}
\begin{figure*}[t]
\centering
\begin{minipage}{0.245\textwidth}\centering
\includegraphics[width=\linewidth]{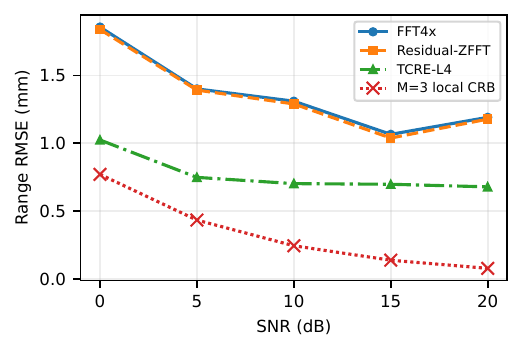}\\[-1mm]{\scriptsize (a) Range RMSE.}
\end{minipage}\hfill
\begin{minipage}{0.245\textwidth}\centering
\includegraphics[width=\linewidth]{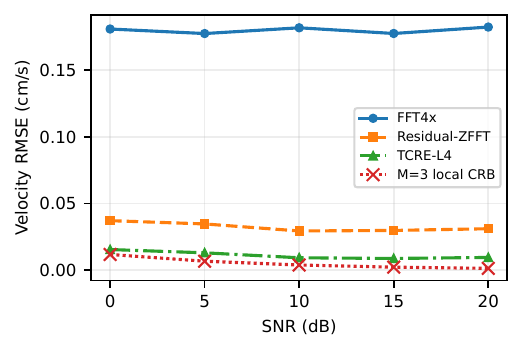}\\[-1mm]{\scriptsize (b) Velocity RMSE.}
\end{minipage}\hfill
\begin{minipage}{0.245\textwidth}\centering
\includegraphics[width=\linewidth]{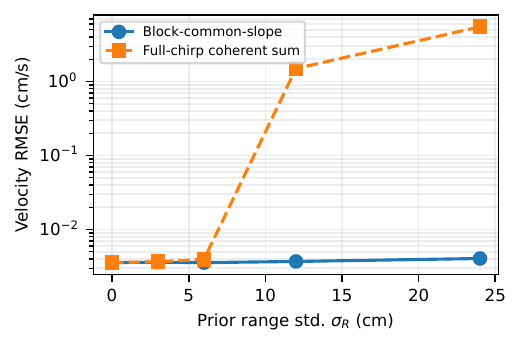}\\[-1mm]{\scriptsize (c) Doppler-stage ablation.}
\end{minipage}\hfill
\begin{minipage}{0.245\textwidth}\centering
\includegraphics[width=\linewidth]{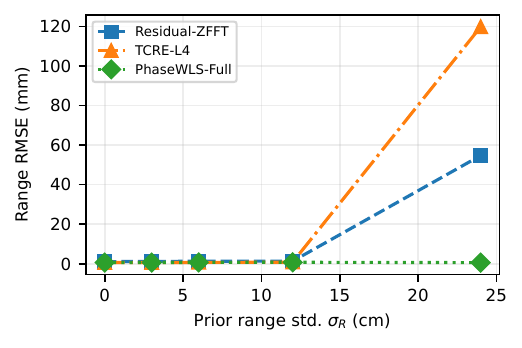}\\[-1mm]{\scriptsize (d) Prior-error robustness.}
\end{minipage}
\caption{Controlled known-truth simulation. \textbf{(a,b)} Range and velocity RMSE for FFT4x, same-prior Residual-ZFFT, TCRE-L4, and the ideal local CRB. \textbf{(c)} Common-slope versus full-chirp Doppler aggregation. \textbf{(d)} Prior-error robustness.}
\label{fig:sim}
\end{figure*}
The experiments examine sub-bin estimation accuracy, information loss
from coherent summarization, and the predicted phase-unwrapping
boundary. Unlike our measured occupancy study \cite{trinh_rasso},
these questions require exact target state and controlled prediction
error. Therefore, we then use known-truth simulation with the 60-GHz configuration as shown in Table~\ref{tab:setup}. The target range and velocity are
drawn as $R\sim\mathcal U(0.8,4.5)$ m and
$v\sim\mathcal U(-0.75,0.75)$ m/s, and the azimuth varies over
$\pm25^\circ$ only to vary receiver phase 
. Two stationary nuisance
reflectors are fixed at
$(1.35\,\mathrm{m},-35^\circ,0.12)$ and
$(3.85\,\mathrm{m},30^\circ,0.08)$, where the last entry is
relative amplitude. The target initial phases are randomized, and no minimum target--clutter range separation is imposed. The  default prior errors are
$e_R\sim\mathcal N(0,0.03^2)$ m and
$e_v\sim\mathcal N(0,0.06^2)$ m/s.
Although Eq.~(\ref{eq:model_coupled}) is used for the derivation, numerical truth is generated from the exact delayed-chirp constant-velocity model 
\begin{equation}
    x_m^{\rm ex}[q,n]\propto e^{-j2\pi\Psi(q,n)}
\end{equation}
with $\Psi(q,n)=f_c\tau+St_n\tau-\tfrac12S\tau^2$ and $\tau(q,n)=\frac{2[R+v(\tau_q+t_n)]}{c}$.
At $v=0.75$ m/s, the target moves \MotionDistanceMaxMm\,mm between the first and last chirp starts. Figure~\ref{fig:blocktrade}(right) evaluates the effect of this motion by comparing the separable and coupling-aware estimators as radial velocity increases. All subsequent TCRE experiments use the coupling-aware processing and exact delayed-chirp signal model above.
We choose spectral baselines: a strengthened conventional
range--Doppler processor, denoted FFT4x and residual zoom FFT, denoted as Residual-ZFFT. FFT4x applies fast/slow-time Hann windows,
$4\times$ zero-padding ($256\times512$), noncoherent three-RX
power summation, three-point quadratic log-power interpolation, and a
0.5--5.0 m/$\pm1.2$ m/s global search. Residual-ZFFT receives the
same $(R^-,v^-)$ and coupling-aware derotation as TCRE. It then uses
4096-point Hann-windowed Doppler and range FFTs with receiver-power
summation and the same interpolation rule, searching the full
unambiguous residual-Doppler interval and
$|\Delta f_b|\le1/(2PT_s)=31.25$ kHz. PhaseWLS-Full applies the
same phase estimator as TCRE to all $N=64$ fast-time samples but without
block compression. Intuitively, FFT4x tests global versus tracked processing,
Residual-ZFFT provides the same-prior spectral comparison, and
PhaseWLS-Full isolates the effect of coherent summarization. The main SNR, prior-error, and sequential experiments include the two nuisance reflectors. CRB, block-count, common-slope, moving-model, and capture tests omit clutter to isolate the corresponding mechanism. The main comparison uses 1,000 matched scenes per SNR (5,000 total). The remaining tests use the verified trial counts reported with their corresponding figures.
\section{Results and Discussion}
\label{sec:results}
\subsection{Representation and estimation accuracy}
We first check the assumptions used to choose the tracked representation. 
Figure~\ref{fig:blocktrade}(left) shows the 10-dB range RMSE of the $L=4$ operating point is only \BlockEmpiricalPenaltyPct\% above $L=32$, close to the \BlockTheoryPenaltyPct\% ideal
standard-deviation penalty, while reducing retained fast-time data by $16\times$. 
Figure~\ref{fig:blocktrade}(right) validates the moving-target model. At $v=0.75$ m/s, the separable approximation gives
\MotionNarrowVel\,cm/s velocity RMSE, whereas the coupling-aware estimator gives \MotionCoupledVel\,cm/s and its corresponding range RMSE is \MotionCoupledRange\,mm. The separate three-receiver 10-dB experiment gives CRBs of \CRBRangeTen\,mm and \CRBVelTen\,cm/s, compared with \TCREIdealRangeTen\,mm and \TCREIdealVelTen\,cm/s over 5,000 exact-signal trials experiment. In the matched cluttered study shown in Fig.~\ref{fig:sim}(a,b), TCRE gives \TCRETenRange\,mm range RMSE and \TCRETenVel\,cm/s velocity RMSE at 10 dB, compared with \ZFFTTenRange\,mm and \ZFFTTenVel\,cm/s for same-prior Residual-ZFFT. The corresponding reductions are \RangeRedTen\% (95\% CI:
\RangeCILowTen--\RangeCIHighTen\%) and \VelRedTen\%
(\VelCILowTen--\VelCIHighTen\%). Across 0--20 dB range, they remain
\RangeRedMin--\RangeRedMax\% and \VelRedMin--\VelRedMax\%.
PhaseWLS-Full gives \FullPhaseTenRange\,mm range RMSE at 10 dB,
showing that the ZFFT gap is an end-to-end estimator comparison rather
than a compression gain.
\subsection{Local robustness and sequential use}
Figure~\ref{fig:sim}(c) validates the common-slope stage: at $\sigma_R=12$ cm and $\sigma_v=24$ cm/s it gives \CommonSlopeTwelve\,cm/s velocity RMSE versus
\FullChirpTwelve\,cm/s after full-chirp coherent summation, a \CommonSlopeRatioTwelve$\times$ separation. Figure~\ref{fig:sim}(d) shows the complementary prior-error behavior: TCRE and the matched-span ZFFT eventually leave their local operating regions, whereas the uncompressed phase control retains a wider sample-level phase branch.
\begin{figure}[t]
\centering
\includegraphics[width=.64\columnwidth]{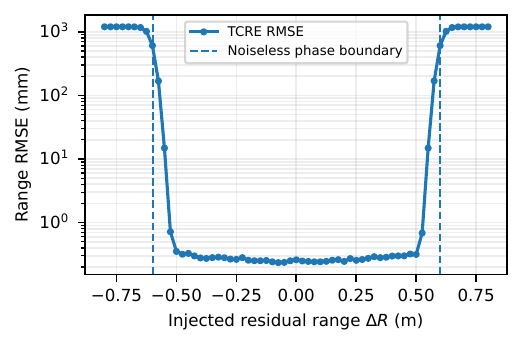}
\caption{Deterministic $L=4$ capture test at 10 dB
(400 trials/point). Dashed lines mark the predicted
$\pm0.60$-m boundary.}
\label{fig:capturetest}
\end{figure}
The deterministic sweep in Fig.~\ref{fig:capturetest} directly tests Eq.~\eqref{eq:capture_corrected}. For $\Delta f_D=0$, the predicted
noiseless boundary is $|\Delta R|=\CaptureNoiselessM$ m. At 10 dB,
RMSE remains below 1 mm through the sampled
$|\Delta R|=\CaptureSubmmSampledM$ m point, then rises to
14.84 mm at 0.55 m and 169.1 mm at 0.575 m. At the sampled
$\pm0.600$-m points, branch-failure rates are 73.0--75.25\%, with
all trials failing for $|\Delta R|\ge0.625$ m. Thus, the analytical boundary describes the noiseless phase branch limit, not guaranteed
sub-millimetric accuracy at its edge. Finally, we close the prediction loop on 80 smooth radial trajectories of 150 frames at $\Delta T=0.1$ s. The trajectories follow
$R_k=2.65+A\sin(2\pi f k\Delta T+\psi)$ with
$A\sim\mathcal U(0.4,0.9)$ m and
$f\sim\mathcal U(0.03,0.08)$ Hz. 
Frame 0 uses FFT4x, and  $R_k^-=\hat R_{k-1}+\hat v_{k-1}\Delta T$ and $v_k^-=\hat v_{k-1}$ with no resets. Across
\SequentialUpdates\ local updates, we observed \SequentialBoundaryViolations\ pre-estimation boundary violations, \SequentialBranchFailures\ branch failures, and \SequentialTrackLosses\ track losses. The largest prediction errors are  \SequentialMaxPredRangeMm\,mm and \SequentialMaxPredVelCms\,cm/s. Thus, the result demonstrates
stability for these smooth single-target dynamics and Fig.~\ref{fig:capturetest} separately probes the capture boundary.

\section{Conclusion}
\label{sec:conclusion}
This paper presented TCRE,
which uses an existing FMCW track before spectral formation and refines
residual range and velocity from four coherent sums per chirp. With
$L=4$, it retains 93.77\% of zero-residual fast-time information at
$16\times$ representation reduction and lowers range/velocity RMSE by
\RangeRedMin--\RangeRedMax\% and
\VelRedMin--\VelRedMax\% relative to same-prior Residual-ZFFT.
The method is intentionally local and future work will address multi-target and hardware operation.
\clearpage
\balance
\bibliographystyle{IEEEbib}
\bibliography{refs}
\end{document}